\documentclass[a4paper,final,11pt,onecolumn]{article}%
\usepackage{amsfonts}
\usepackage{amsmath}
\usepackage{amssymb}
\usepackage{graphicx}
\usepackage{version}
\usepackage{accents}%
\providecommand{\U}[1]{\protect\rule{.1in}{.1in}}
\begin{document}

\title{The\textit{ }integration constants method in quantum field theory}
\author{{\small Z. Belhadi\thanks{Email : zahir.belhadi@univ-bejaia.dz}}\\{\small Laboratoire de physique th\'{e}orique, Facult\'{e} des sciences
exactes.}\\{\small Universit\'{e} de Bejaia. 06000 Bejaia, Alg\'{e}rie.}}
\date{}
\maketitle

\subsection*{Abstract :}

\ \ \ \ \ \ Recently, Belhadi and al.\cite{nous} developed a new approach to
quantize classical soluble systems based on the calculation of brackets among
fundamental variables using the constants of integration (CI method). In this
paper, we will apply this approach in some exactly soluble constrained
Hamiltonian systems. We will complete our work with some applications in
quantum field theory : Majorana neutrino,\ the scalar field on the light-cone,
the O(2) nonlinear sigma model and the chiral boson theory.

\textbf{Key words} : Quantization, constrainted systems, field theory, soluble
systems, integration constants.

\section{Introduction}

\qquad It is well known in quantum mechanics that the quantization of
classical systems can be done in two principal ways: the first one is the
Feynman path integrals where starting from a classical Lagrangian, the sum
over all possible trajectories is performed to obtain the probability
amplitude. The second one is the canonical quantization, which uses the
Hamiltonian formulation and the Poisson brackets to determine the expressions
of the different commutators of the operators corresponding to the classical
variables and the wave function. However, in practice, these methods are not
easy to apply due to certain physical requirements that any consistent theory
must satisfy. For example, the problem of operators ordering, the difficulties
related to gauge symmetries, and the constraints that appear when the
Lagrangian is singular.

In the case of constrained Hamiltonian systems, the works of Dirac and
Bergmann on this topic is are very fundamental tool \cite{Dirac}. Indeed, they
developed a generalized Hamiltonian formalism permitting the canonical
quantization of these systems with singular Lagrangians using the Dirac
brackets which replace Poisson brackets in the presence of constraints. Dirac
went further when he showed the link between these constraints and the gauge
symmetry of some types of singular Lagrangians when he defined first and
second class constraints. In 1988, another approach emerged from the work of
Faddeev and Jackiw on singular systems \cite{FJ}. They developed a much more
straightforward symplectic method without distinction between the constraints
of first and second class and they get the same results as Dirac.

Recently, developing the idea that the general solution of the classical
motion equations of any system contains implicitly the brackets necessary for
its quantization, Belhadi and al. constructed a new approach called the
integration constants method (CI) using this solution without the help of
Dirac formalism or FJ\ method \cite{nous}. Indeed, it is possible to use the
integration constants to determine the different brackets among the
fundamental variables without mentioning anything about constraints. After
making comparison with Dirac and FJ procedures in several examples, we have
successfully applied our approach in the case of Dirac field \cite{nous}. it
is noteworthy that our approach coincides with a requirement imposed by Pauli
in the case of free relativistic field theory \cite{P}.

In the following, first, we recall briefly the principle of the CI method in
order to apply it in specific cases. Then, a special study is dedicated to the
nonlinear sigma model and Majorana particle to show that the CI method works
very well in quantum field theory. We finish our work with the light-cone free
scalar field and chiral boson quantization.

\section{Method of integration constants (CI)}

\qquad Consider a classical system described by a singular autonomous
Lagrangian $L(q,\dot{q})$ where $q=(q_{1},...,q_{N})$ are the generalized
coordinates and $\dot{q}=(\dot{q}_{1},...,\dot{q}_{N})$ the generalized
velocities. Suppose we have the (general) analytical solutions $q(t)=\tilde
{q}(t,C)$ of Euler-Lagrange equations and the momenta$\ p(t)=\tilde{p}(t,C)$
$\left(  p_{i}=\frac{\partial L}{\partial\dot{x}}\right)  $, where
$C=(C_{1},C_{2},...,C_{M})$ is the set of the \underline{independent}
integration constants. For constrained\textbf{\footnote{$M=2N$ corresponds to
an unconstrained system. For the constrained systems, each constraint
eliminates one variable.}} systems we have obviously $M<2N.$

In the case when the general solution depends on arbitrary functions due to
the gauge symmetry, we must first choose these functions by adding new
conditions (fixing the gauge) before going to the canonical formalism and
defining any bracket.

From the analytical solutions of the equations of motion, we can write the
Hamiltonian\footnote{We can also obtain the Hamiltonian by putting the
solutions into the Legendre transformation $H=\sum_{i}\frac{d\tilde{q}%
_{i}(t,C)}{dt}\tilde{p}_{i}(t,C)-L\left(  \tilde{q}_{i}(t,C),\frac{d\tilde
{q}_{i}(t,C)}{dt}\right)  .$ In this way we do not have to inverse the momenta
with respect to the velocities.} as $H(q(t),p(t))=H(\tilde{q}(t,C),\tilde
{p}(t,C))=\tilde{H}(C).$ In CI method \cite{nous}, we determine the
integration constants brackets with the help of the following property%
\begin{equation}
\left\{
\begin{array}
[c]{c}%
\frac{\partial}{\partial t}\tilde{q}_{i}(t,C)=%
%TCIMACRO{\tsum \limits_{k,l=1}^{M}}%
%BeginExpansion
{\textstyle\sum\limits_{k,l=1}^{M}}
%EndExpansion
\{C_{k},C_{l}\}\frac{\partial\tilde{q}_{i}}{\partial C_{k}}\frac
{\partial\tilde{H}}{\partial C_{l}}\ \ \ \ i=1...N\\
\frac{\partial}{\partial t}\tilde{p}_{i}(t,C)=%
%TCIMACRO{\tsum \limits_{k,l=1}^{M}}%
%BeginExpansion
{\textstyle\sum\limits_{k,l=1}^{M}}
%EndExpansion
\{C_{k},C_{l}\}\frac{\partial\tilde{p}_{i}}{\partial C_{k}}\frac
{\partial\tilde{H}}{\partial C_{l}}\ \ \ \ i=1...N
\end{array}
\right.  \label{hyhyhy}%
\end{equation}
These $2N$ equations contain $M(M-1)/2$ unknown brackets $\{C_{k},C_{l}\},$
with $k,l=1...M$. Our method involves the determination of the brackets
$\{C_{k},C_{l}\}$ via a simple identification. But, sometimes, we have to add
supplementary terms to the Lagrangian like $\eta q_{i}$ and redo all the
calculations and put $\eta=0$ at the end \cite{nous}$.$ Using the brackets
$\{C_{k},C_{l}\},$ we can compute the brackets $\left\{  q_{i},q_{j}\right\}
$, $\left\{  p_{i},p_{j}\right\}  $ and $\left\{  q_{i},p_{j}\right\}  $ more
easily than with any other existing approaches$.$ If the result of the
calculation depends on the integration constants, it is possible to make them
disappear by inversing the solution $\tilde{q}(t,C)$ and $\tilde{p}(t,C)$. In
deed, once we replaced the fundamental variables\textbf{ }$q_{i}$\textbf{ }and
\textbf{ }$p_{i}$\textbf{ }by the solution $\tilde{q}_{i}(t,C)$\textbf{
}and\textbf{ }$\tilde{p}_{i}(t,C),$\textbf{ }we obtain the property%
\begin{equation}
\left\{  f,g\right\}  =\sum_{k,l=1}^{M}\left\{  C_{k},C_{l}\right\}
\frac{\partial\tilde{f}}{\partial C_{k}}\frac{\partial\tilde{g}}{\partial
C_{l}} \label{EqFondamental}%
\end{equation}

We can see that in our method, we do not even talk about any constraints and
any classifications unlike other approaches.

As an illustration, let's start with this two examples :

\textbf{1}\textit{.} The Christ--Lee model \cite{g} is described by the
singular Langrangian $L=\frac{1}{2}{\Large (}\dot{r}^{2}+r^{2}(\dot{\theta
}-z)^{2}{\Large )}-V(r),$ where $r$ and $\theta$ are plane polar coordinates,
$z$ is another generalized coordinate, and $V(r)$ is a potential. In the case
where $V(r)=\frac{1}{2}r^{2},$ Euler-Langange equations are $\ddot{r}%
=r(\dot{\theta}-z)^{2}-r,$\ $\frac{d}{dt}(r^{2}(\dot{\theta}-z))=0$
and\ $0=-r^{2}(\dot{\theta}-z)\ $and the general solution is $r=a\cos
(t)+b\sin(t),$ $\theta(t)=\varepsilon(t)$ and $z(t)=\dot{\varepsilon}(t)$
where $a$ and $b$ are the only integration constantes and $\varepsilon(t)$ an
arbitrary time function. In order to fix the gauge, we choose $\varepsilon
(t)=0$ and the solution will be simplified to $r(t)=a\cos(t)+b\sin(t),$
$\theta(t)=0$ and $z(t)=0.$

We deduce that $p_{r}(t)=-a\sin(t)+b\cos(t),$ $p_{\theta}(t)=0$ and
$p_{z}(t)=0,$ and the Hamiltonian can be expressed as $H=\dot{r}p_{r}%
+\dot{\theta}p_{\theta}+\dot{z}p_{z}-L=1/2(a^{2}+b^{2}).$ Using the equation
$\dot{r}=\{r,H\},$ we obtain
\[
-a\sin(t)+b\cos(t)=b\cos(t)\{a,b\}+a\sin(t)\{b,a\}.
\]
We deduce by identification the integration constants bracket $\{a,b\}=1.$ The
calculus of the fundamental brackets gives $\{r,p_{r}\}=1$ when the others vanish.

\textbf{2}\textit{.} Now, let's study the case of the fermionic harmonic
oscillator \cite{bbbb} described by the Lagrangian $L=\frac{i}{2}\left(
\bar{\psi}\dot{\psi}-\overset{\cdot}{\bar{\psi}}\psi\right)  -\omega\bar{\psi
}\psi$ where $\bar{\psi}$ and $\psi$ and are two Grassmann independent odd
variables. Then, equations of motion are $\overset{\cdot}{\bar{\psi}}%
=i\omega\bar{\psi}$ and $\dot{\psi}=-i\omega\psi$ whose the solution is
$\psi=ae^{-i\omega t}$ and $\bar{\psi}=\bar{a}e^{i\omega t}$ whereas the
canonical hamiltonian is $H=\omega\bar{\psi}\psi=\omega\bar{a}a$ ($\bar{a}$
and $a$ are two Grassmann integration constants)$.$ At this stage, the
Hamilton equation $\dot{\psi}=\{\psi,H\}$ implies that $-ia=\{a,\omega\bar
{a}a\},$ therefore
\[
-ia=\{a,\bar{a}a\}=\{a,\bar{a}\}a+(-1)^{\epsilon_{a}\epsilon_{\bar{a}}}\bar
{a}\{a,a\}
\]
where $\epsilon_{a}=1$ and $\epsilon_{\bar{a}}=1$ are the parities of $a$ and
$\bar{a}$ in this order. Because we are working with Grassmann variables, we
assume that the previous bracket $\{$ $;$ $\}$ has the same propreties as the
anticommuting variables Poisson brackets \cite{bbbbb}$.$ In other words,
$-ia=\{a,\bar{a}\}a-\bar{a}\{a,a\}$ and by a simple identification, we deduce
the brackets $\{a,\bar{a}\}=-i$ and $\{a,a\}=0.$ We can obtain the brackets
$\{\bar{a},a\}=-i$ and $\{\bar{a},\bar{a}\}=0$ using $\overset{\cdot}%
{\bar{\psi}}=\{\bar{\psi},H\}.$ Knowing that $\Pi_{\psi}=\frac{\partial
L}{\partial\dot{\psi}}=-\frac{i}{2}\bar{\psi}$ and $\Pi_{\bar{\psi}}%
=\frac{\partial L}{\partial\overset{\cdot}{\bar{\psi}}}=-\frac{i}{2}\psi,$ the
only non vanishing brackets with our Lagrangian are $\{\psi,\bar{\psi}\}=-i$
and $\{\psi,\Pi_{\psi}\}=\{\bar{\psi},\Pi_{\bar{\psi}}\}=2.$

\section{Application in field theory}

\qquad In this section, we consider our approach in quantum framework to find
the fundamental commutation relations already known in quantum field theory.

\subsection{O(2) Nonlinear sigma model}

\qquad The nonlinear sigma model was introduced by Gell-Mann and L\'{e}vy in
1960 \cite{gl} to describe interaction implemented within constraint between
the fields. It is a (1+1) dimension non-trivially solvable system in quantum
field theory. More precisely, the O(2) nonlinear sigma model in one space
($x$) one time ($t$) dimension is described by the singular Lagrangian density
$\mathcal{L}=\frac{1}{2}\partial_{\mu}\phi_{a}^{{}}$ $\partial^{\mu}\phi
_{a}^{{}}+\lambda\left(  \phi_{a}\phi_{a}-1\right)  $ where $a\in\left\{
1,2\right\}  $ and $\mu\in\left\{  0,1\right\}  .$ Euler-Lagrange equations
for the real fields $\lambda$ and $\phi_{a}^{{}}$ are $\phi_{a}\phi_{a}-1=0$
and $\partial_{t}^{2}\phi_{a}-\partial_{x}^{2}\phi_{a}=2\lambda\phi_{a}.$
Let's put $\phi_{1}=\varphi$ and $\phi_{2}=\psi$ in order to write the
previous equations in the form $\varphi^{2}+\psi^{2}=1$ and\ $\square
\varphi=2\lambda\varphi$ and\ $\square\psi=2\lambda\psi.$ We deduce that%
\begin{equation}
\varphi^{2}+\psi^{2}=1\ \ \ \ \ \ \ \ \psi\square\varphi-\varphi\square\psi=0
\label{mm1}%
\end{equation}
The first equation above suggests puting $\varphi=\cos(\theta)$ and $\psi
=\sin(\theta)$ and after replacing them in (\ref{mm1}), we obtain the equation
$\square\theta=0.$ This is D'Alembert equation which is also Klein-Gordon
equation of spinless and massless particle without charge. It's well-known
that its classical general solution in two-dimensional $(x,t)$ is%
\begin{equation}
\tilde{\theta}(t,x)=\int dk_{1}{\large (}f_{k}(t,x)a(k_{1})+f_{k}^{\ast
}(t,x)a^{\ast}(k_{1}){\large )} \label{mm2}%
\end{equation}
where $f_{k}=\frac{e^{-i(k_{0}t-k_{1}x)}}{\sqrt{(2\pi)\text{ }2k_{0}}}$ and
$k_{0}=|k_{1}|$. The $a(k_{1})$ and $a^{\ast}(k_{1})$ are arbitrary functions
of $k_{1}$ (integration constants). Then, the general solution of (\ref{mm1})
will be $\varphi(t,x)=\cos\tilde{\theta}(t,x)$ and $\ \psi(t,x)=\sin
\tilde{\theta}(t,x).$ Knowing that $\Pi_{1}=\Pi_{\varphi}=\frac{\partial
L}{\partial\dot{\varphi}}=\dot{\varphi}$ and $\Pi_{2}=\Pi_{\psi}%
=\frac{\partial L}{\partial\dot{\psi}}=\dot{\psi},$ we can express the
Hamiltonian $H=\int dx\left(  \Pi_{\varphi}\dot{\varphi}+\Pi_{\psi}\dot{\psi
}-\mathcal{L}\right)  $\ as%
\begin{equation}
H=\int dk_{1}k_{0}a^{\ast}(k_{1})a(k_{1}) \label{mm4}%
\end{equation}
With the help of Hamilton equation $\dot{\varphi}=$ $\left\{  \varphi
,H\right\}  ,$we obtain\footnote{Here we used the relation $\left\{
h(g),f\right\}  =\frac{\partial h}{\partial g}\left\{  g,f\right\}  $ which
can be verified with the help of propriety $\left\{  g^{n},f\right\}
=ng^{n-1}\left\{  g,f\right\}  $ and the fact that $h(g)=\sum_{n=0}^{\infty
}a_{n}g^{n}$.} the relation below $\overset{\boldsymbol{\cdot}}{\tilde{\theta
}}=\left\{  \tilde{\theta},H\right\}  .$ Using (\ref{mm2}), (\ref{mm4}) and
the above equation, we deduce by identification
\begin{align}
\left\{  a(k_{1}^{{}}),a^{\ast}(k_{1}^{\prime})a(k_{1}^{\prime})\right\}   &
=-ia(k_{1}^{\prime})\delta(k_{1}-k_{1}^{\prime})\label{951}\\
\left\{  a^{\ast}(k_{1}),a^{\ast}(k_{1}^{\prime})a(k_{1}^{\prime})\right\}
&  =ia^{\ast}(k_{1}^{\prime})\delta(k_{1}-k_{1}^{\prime})
\end{align}
Using Liebniz rule
\begin{align}
&  a^{\ast}(k_{1}^{\prime})\left\{  a(k_{1}^{{}}),a(k_{1}^{\prime})\right\}
+\left\{  a(k_{1}^{{}}),a^{\ast}(k_{1}^{\prime})\right\}  a(k_{1}^{\prime
})\nonumber\\
&  =-ia(k_{1}^{\prime})\delta(k_{1}-k_{1}^{\prime})
\end{align}
and%
\begin{align}
&  a^{\ast}(k_{1}^{\prime})\left\{  a^{\ast}(k_{1}),a(k_{1}^{\prime})\right\}
+\left\{  a^{\ast}(k_{1}),a^{\ast}(k_{1}^{\prime})\right\}  a(k_{1}^{\prime
})\nonumber\\
&  =ia^{\ast}(k_{1}^{\prime})\delta(k_{1}-k_{1}^{\prime})
\end{align}
Finally, by direct identification we obtain the following brackets%
\begin{align}
\left\{  a(k_{1}^{{}}),a^{\ast}(k_{1}^{\prime})\right\}   &  =-i\delta
(k_{1}-k_{1}^{\prime})\\
\left\{  a(k_{1}^{{}}),a(k_{1}^{\prime})\right\}   &  =\left\{  a^{\ast}%
(k_{1}),a^{\ast}(k_{1}^{\prime})\right\}  =0
\end{align}

At this stage, it's convenient to calculate the bracket $\left\{
\tilde{\theta}(t,x),\overset{\boldsymbol{\cdot}}{\tilde{\theta}}(t,x^{\prime
})\right\}  $ using (\ref{mm2}). Indeed, we obtain the bracket below%

\begin{equation}
\left\{  \tilde{\theta}(t,x),\overset{\boldsymbol{\cdot}}{\tilde{\theta}%
}(t,x^{\prime})\right\}  =\delta(x-x^{\prime})
\end{equation}
From the relations $\varphi(t,x)=\cos\tilde{\theta}(t,x)$ and $\ \psi
(t,x)=\sin\tilde{\theta}(t,x)$, we deduce the following brackets%
\begin{equation}%
\begin{array}
[c]{l}%
\{\phi_{a},\phi_{b}\}=0\\
\{\phi_{a},\Pi_{b}\}=\left(  \delta_{ab}-\frac{\phi_{a}\phi_{b}}{\phi_{c}^{2}%
}\right)  \delta(x-x^{\prime})\\
\{\Pi_{a},\Pi_{b}\}=-\frac{1}{\phi_{c}^{2}}(\phi_{a}\Pi_{b}-\phi_{b}\Pi
_{a})\delta(x-x^{\prime})
\end{array}
\end{equation}
\qquad\ \ \ Thus, by the mean of the CI method, we derived the necessary
brackets to quantize the nonlinear sigma model. The same result is obtained in
\cite{mk} using Dirac formalism and in \cite{ijtp} with Faddeev-Jackiw
approach. This relations are still valid in the case of O(N) nonlinear sigma
model, that strongly indicates the consistency our method even for nonlinear situations.

\subsection{Application to Majorana neutrino}

\qquad In 1937, Etore Majorana introduced his particle as a neutral fermion
with spin 1/2, which is its own antiparticle, but his work was considered just
as an other mathematical formulation of Dirac equation until the beginning of
1960's, when physicists started to ask the question without answer even now :
the neutrinos are Dirac fermions or Majorana fermions ?

As an application of our method, let's study the case of a left-handed
Majorana field, whose the Lagrangian density has the form
\begin{equation}
\mathcal{L}=\eta^{\dag}i\bar{\sigma}^{\mu}\partial_{\mu}\eta-\frac{m}%
{2}\left(  \eta^{T}i\sigma^{2}\eta-\eta^{\dag}i\sigma^{2}\eta^{\ast}\right)
\end{equation}
where $\bar{\sigma}^{\mu}=(\sigma^{0},-\sigma^{1},-\sigma^{2},-\sigma^{3})$
and $\eta$ is a two components complex field\footnote{$\bar{\sigma}^{\mu
}\partial_{\mu}=\partial_{t}-\vec{\sigma}.\vec{\nabla}$ where $\mathbf{\sigma
}=(\sigma^{1},\sigma^{2},\sigma^{3})$ are the Pauli matrices}. The
Euler-Lagrange equations lead to Majorana equation $\bar{\sigma}^{\mu}%
\partial_{\mu}\eta+m\sigma^{2}\eta^{\ast}$ $=0$ whose general solution is
\cite{Masaru}
\begin{align}
\eta &  =\int\sum_{s=1}^{2}d\vec{k}\left(  a_{s}(\vec{k})w_{s}%
^{\text{{\tiny (1)}}}(k)f_{k}(x)\right. \nonumber\\
&  \text{ \ \ \ \ \ \ \ \ \ \ \ \ \ }\left.  +a_{s}^{\dagger}(\vec{k}%
)v_{s}w_{s}^{\text{{\tiny (2)}}}(k)f_{k}^{\ast}(x)\right)  \label{qirdus}%
\end{align}
where $f_{k}(x)=\sqrt{\frac{m}{(2\pi)^{3}k_{0}}}e^{-ikx}$, $k_{0}=\sqrt
{\vec{k}^{2}+m^{2}}$ and $a_{s}(\vec{k})$ and $a_{s}^{\dagger}(\vec{k})$ are
operators. $w_{s}^{\text{{\tiny (1)}}}(k)$ and $w_{s}^{\text{{\tiny (2)}}}(k)$
are two spinors which can be expressed using the spinors $\chi_{1}=(1$
\ $0)^{T}$ and $\chi_{2}=(0$ \ $1)^{T}$as $w_{s}^{\text{{\tiny (1)}}}%
(k)=\sqrt{\frac{k_{0}+m}{2m}}\left(  1-\frac{\vec{\sigma}\mathbf{.}\vec{k}%
}{k_{0}+m}\right)  \frac{1}{\sqrt{2}}\chi_{s}$\ and

\noindent$w_{s}^{\text{{\tiny (2)}}}(k)=\sqrt{\frac{k_{0}+m}{2m}}\left(
1-\frac{\vec{\sigma}\mathbf{.}\vec{k}}{k_{0}+m}\right)  \frac{i\sigma^{2}%
}{\sqrt{2}}\chi_{s}$. Using the solution (\ref{qirdus}), the Hamiltonian
$H=\frac{1}{2}\int d\vec{x}i(\eta^{\dag}\partial_{t}\eta-\partial_{t}%
\eta^{\dag}\eta)$ takes the form $H=\int d\vec{k}k_{0}\sum_{s=1}^{2}N_{s}%
(\vec{k})$ where

\noindent\ $N_{s}(\vec{k})=\frac{1}{2}\left(  a_{s}^{\dag}(\vec{k})a_{s}%
(\vec{k})-a_{s}(\vec{k})a_{s}^{\dag}(\vec{k})\right)  .$

Let's now apply our approach in the quantum context using the Heisenberg
equation $\dot{\eta}=\frac{1}{i}\left[  \eta,H\right]  .$ Directly, we read
off the following brackets
\begin{equation}
\left[  a_{s}(\vec{k}),N_{s^{\prime}}(\vec{k}^{\prime})\right]  =a_{s^{\prime
}}(\vec{k}^{\prime})\delta_{ss^{\prime}}\delta(\vec{k}-\vec{k}^{\prime})
\label{azertyuio}%
\end{equation}%
\begin{equation}
\left[  a_{s}^{\dagger}(\vec{k}),N_{s^{\prime}}(\vec{k}^{\prime})\right]
=-a_{s^{\prime}}^{\dagger}(\vec{k}^{\prime})\delta_{ss^{\prime}}\delta(\vec
{k}-\vec{k}^{\prime}). \label{azertyuiop}%
\end{equation}
For fermions, we use the propriety $[A,BC]=-B\left[  A,C\right]  _{+}+\left[
A,B\right]  _{+}C$ in order to express these commutators by the means of
anticommutators%
\begin{align}
\left[  a_{s}(\vec{k}),N_{s^{\prime}}(\vec{k}^{\prime})\right]   &  =-\frac
{1}{2}a_{s^{\prime}}^{\dag}(\vec{k}^{\prime})\left[  a_{s}(\vec{k}%
),a_{s^{\prime}}(\vec{k}^{\prime})\right]  _{+}\nonumber\\
&  +\frac{1}{2}\left[  a_{s}(\vec{k}),a_{s^{\prime}}^{\dag}(\vec{k}^{\prime
})\right]  _{+}a_{s^{\prime}}(\vec{k}^{\prime})\nonumber\\
&  +\frac{1}{2}a_{s^{\prime}}(\vec{k}^{\prime})\left[  a_{s}(\vec
{k}),a_{s^{\prime}}^{\dag}(\vec{k}^{\prime})\right]  _{+}\nonumber\\
&  -\frac{1}{2}\left[  a_{s}(\vec{k}),a_{s^{\prime}}(\vec{k}^{\prime})\right]
_{+}a_{s^{\prime}}^{\dag}(\vec{k}^{\prime}) \label{klm}%
\end{align}
Then by identification of (\ref{azertyuio}) with (\ref{klm}), we obtain the
anticommutation rules%
\begin{align}
\left[  a_{s}(\vec{k}),a_{s^{\prime}}^{\dag}(\vec{k}^{\prime})\right]  _{+}
&  =\delta_{ss^{\prime}}\delta(\vec{k}-\vec{k}^{\prime})\\
\left[  a_{s}(\vec{k}),a_{s^{\prime}}(\vec{k}^{\prime})\right]  _{+}  &  =0
\end{align}
Now, from equation (\ref{azertyuiop}), we deduce the anticommutators%
\begin{align*}
\left[  a_{s}^{\dagger}(\vec{k}),a_{s^{\prime}}(\vec{k}^{\prime})\right]
_{+}  &  =\delta_{ss^{\prime}}\delta(\vec{k}-\vec{k}^{\prime})\\
\left[  a_{s}^{\dagger}(\vec{k}),a_{s^{\prime}}^{\dagger}(\vec{k}^{\prime
})\right]  _{+}  &  =0
\end{align*}
These rules of quantization are exactly identical to the results of the
canonical quantization of Majorana field \cite{Masaru,gitman}.

\subsection{The (1+1) light-cone quantization of free scalar field}

\ \ \ \ \ Unlike the conventional quantization where the different brackets
are calculated at the same time $x^{0},$ with the light-cone formulation,
these brackets are evaluated on the line $x^{+}=x^{0}+x^{1}$ considered as
time$.$ Indeed, constraints arise resulting in the Hamiltonian formulation on
the cone of light due to this change of coordinates. Dirac's method \cite{LC2}
, Schwinger's action principle \cite{LC77} and Faddeev-Jackiw's symplectic
approach \cite{LC777} can be employed with success to deduce the correct
canonical equal $x^{+}$ commutation relations nacessary to quantize this type
of systems. Our goal is to show how the CI method can be used to obtain the
correct brackets directly by a simple identification.

The light-cone coordinates are defined as $x^{+}=x^{0}+x^{1}$ and $x^{-}%
=x^{0}-x^{1}$ which can be inversed to obtain $x^{0}=\frac{x^{+}+x^{-}}{2}$
and $x^{1}=\frac{x^{+}-x^{-}}{2}.$ The scalar product becomes $ab=a^{\mu
}b_{\mu}=\frac{1}{2}a^{+}b^{-}+\frac{1}{2}a^{-}b^{+}$ where $\mu=0,1.$ So, our
metric is $g_{+-}=g_{-+}=1/2$ and $g_{++}=g_{--}=0.$ We can deduce the
relations $\ \partial_{0}=\partial_{+}+\partial_{-}$ and $\partial
_{1}=\partial_{+}-\partial_{-}.$

With this change of variables, the free scalar field action $S=\int dx^{0}\int
dx^{1}
\newline\left(  \frac{1}{2}\partial_{\mu}\phi\partial^{\mu}\phi-\frac{m^{2}}%
{2}\phi^{2}\right)  $ can be written as%
\begin{equation}
S=\frac{1}{2}\int{\small dx}^{+}{\small dx}^{-}\left(  2\partial_{+}%
\phi\partial_{-}\phi-\frac{m^{2}}{2}\phi^{2}\right)
\end{equation}
where the factor $\frac{1}{2}$ results from the light-cone Jacobian. From the
Lagrangian density $\mathcal{L=}\partial_{+}\phi\partial_{-}\phi-\frac{m^{2}%
}{4}\phi^{2},$ one cans derive the evolution equation $4\partial_{-}%
\partial_{+}\phi+m^{2}\phi=0,$ whose the general solution is
\begin{equation}
\phi=\int_{0}^{\infty}{\small dk}^{+}\sqrt{\frac{k^{-}}{4\pi m^{2}}%
}{\small (a(k}^{+}{\small )e}^{-ikx}{\small +a}^{\dagger}{\small (k}%
^{+}{\small )e}^{ikx}{\small )} \label{taztaz}%
\end{equation}
where $kx=1/2\left(  k^{+}x^{-}+k^{-}x^{+}\right)  $, $k^{-}=\frac{m^{2}%
}{k^{+}}$ and $a(k^{+})$ and $a_{{}}^{\dagger}(k^{+})$ are operators such that
$a^{\dagger}(k^{+})=a(-k^{+}).$ For this reason, we will work only in the
region where $k^{+}>0$ in order to garantee the independence of the operators
$a(k^{+})$ and $a_{{}}^{\dagger}(k^{+})$. Now, if we substitute the solution
above in the Hamiltonian expression $H=\int{\small dx}^{-}\frac{m^{2}}{4}%
\phi^{2}$, we obtain
\begin{equation}
H=\frac{1}{4}\int_{0}^{\infty}dk^{+}k^{-}\left(  {\small a}^{\dagger
}{\small (k}^{+}{\small )a(k}^{+}{\small )+a(k}^{+}{\small )a}^{\dagger
}{\small (k}^{+}{\small )}\right)  \label{789456123}%
\end{equation}

Now, let's use the Heisenberg equation $\partial_{+}\phi=-i\left[
\phi,H\right]  $ to find the brackets of the operators $a_{{}}(k^{+})$ and
$a_{{}}^{\dagger}(k^{+}).$ In deed, after\ a straightfoward identification, we
get the following relations%
\begin{equation}
\left\{  \hspace{-0.15cm}%
\begin{array}
[c]{c}%
k^{-}a(k^{+})=\frac{1}{2}\int_{0}^{\infty}dk^{\prime+}k^{\prime-}\left[
a(k^{+}),N(k^{\prime+})\right] \\
-k^{-}a^{\dagger}(k^{+})=\frac{1}{2}\int_{0}^{\infty}dk^{\prime+}k^{\prime
-}\left[  a^{\dagger}(k^{+}),N(k^{\prime+})\right]
\end{array}
\right.
\end{equation}
where $N(k^{+})=a^{\dagger}(k^{+})a(k^{+})+a(k^{+})a^{\dagger}(k^{+}).$ Now,
in the same way explained above, one can use (\ref{789456123})\footnote{It
should be noted that $\int_{0}^{\infty}dk^{\prime+}\delta(k^{\prime+}%
-k^{+})f(k^{\prime+})=f(k^{+})$ because $k^{+}\in]0,+\infty\lbrack.$} to
deduce directly the following brackets%
\begin{equation}%
\begin{array}
[c]{l}%
\left[  a(k^{+}),a^{\dagger}(k^{\prime+})\right]  =\delta(k^{+}-k^{\prime+})\\
\left[  a(k^{+}),a(k^{\prime+})\right]  =\left[  a^{\dagger}(k^{+}%
),a^{\dagger}(k^{\prime+})\right]  =0
\end{array}
\end{equation}
These are the correct commutation relations of the creation and the
annihilation operators $a^{\dagger}(k^{+})$ and $a(k^{+})$.

With the help of this resulat, one can evaluate these different equal time
brackets ($x^{+}=x^{\prime+}$):%
\begin{align}
\left[  \phi(x),\phi(x^{\prime})\right]   &  =-\frac{i}{2\pi}\int_{0}%
^{+\infty}\frac{dk^{+}}{k^{+}}\sin(\frac{k^{+}}{2}(x^{-}-x^{\prime
-}))\nonumber\\
&  =-\frac{i}{4}\epsilon(x^{-}-x^{\prime-})
\end{align}%
\begin{align}
\left[  \phi(x),\pi(x^{\prime})\right]   &  =\frac{i}{8\pi}\int_{{}}^{{}%
}dk^{+}e^{\frac{i}{2}k^{+}(x^{-}-x^{\prime-})}\nonumber\\
&  =\frac{i}{2}\delta(x^{-}-x^{\prime-})
\end{align}%
\begin{align}
\left[  \pi(x),\pi(x^{\prime})\right]   &  =\frac{i}{8\pi}\partial_{-}\int
_{{}}^{{}}dk^{+}e^{\frac{i}{2}k^{+}(x^{-}-x^{\prime-})}\nonumber\\
&  =\frac{i}{2}\partial_{-}\delta(x^{-}-x^{\prime-})
\end{align}
Once again, these are exactly the brackets obtained by the other methods cited above.

\subsection{The chiral-boson quantization}

In this section, we propose to quantize the chiral-boson within
the\ Floreanini-Jackiw Lagrangian \cite{jk,cjk}. Indeed, starting with the
Lagrangian density $\mathcal{L=}\frac{1}{2}\left(  \partial_{t}\phi
\partial_{x}\phi-\left(  \partial_{x}\phi\right)  ^{2}\right)  ,$ one cans
derive the motion equation $\partial_{t}\partial_{x}\phi=\partial_{x}^{2}%
\phi,$ whose the general solution is
\begin{equation}
{\small \phi=}\int_{0}^{\infty}\frac{{\small dk}}{\sqrt{2\pi k}}\left(
{\small a(k)e}^{-ik\left(  t+x\right)  }{\small +a}^{\dagger}{\small (k)e}%
^{ik\left(  t+x\right)  }\right)  \label{hjhj}%
\end{equation}
where $a(k)$ and $a_{{}}^{\dagger}(k)$ are operators$.$ After substitution of
this solution in the Hamiltonian $H=\frac{1}{2}\int{\small dx}\left(
\partial_{x}\phi\right)  ^{2}$, we obtain
\begin{equation}
H=\frac{1}{2}\int_{0}^{\infty}dkk\left(  {\small a}^{\dagger}%
{\small (k)a(k)+a(k)a}^{\dagger}{\small (k)}\right)  \label{hjhjhj}%
\end{equation}

The expression of the solution (\ref{hjhj}) and the Hamiltonian (\ref{hjhjhj})
are similar to those of the solution and the Hamiltoinan (\ref{taztaz}%
,\ref{789456123}), and consequently the calculus will practically be the same
as in last subsection. Indeed, after direct calculus, we get the usual
commutator%
\begin{equation}
\left[  a(k),a^{\dagger}(k^{\prime})\right]  =\delta(k-k^{\prime})
\end{equation}
Using this result$,$ one can obtain the equal-time commutator
\begin{equation}
\left[  \phi(x),\phi(x^{\prime})\right]  =-\frac{i}{2}\epsilon(x-x^{\prime})
\end{equation}
We have here the same result as in \cite{jk,cjk}, obtained in a different way.

\section{Conclusion}

In this paper, we studied the canonical quantization of Lagrangian systems
that are classically exactly solvable using the integration constants method
(CI method), where we showed that fundamental commutation relations can be
derived straightforwardly and efficiently. We first, considered the nonlinear
sigma model in the classical context by constructing the fundamental brackets.
Then, we focused on canonical quantization of Majorana field describing
neutrinos and the free scalar field on the cone of light within the CI method
by establishing repectively, the anticommutation and commutation relations for
creation and destruction operators. In the end, we studied successfully the
case of chiral bonson theory.

A striking result, in all cases, is that one can easily obtain the correct
commutation relations between the operators of creation and annihilation
before knowing anything about the brackets among the fields and their
conjugate momenta. All these verifications and the successes of the CI method
show that this technique is efficient to quantize the exactly solvable systems
by exploiting the integration constants of the general solution obtained from
the equations of motion.

\end{document}